# Joint Mobile IAB Node Positioning and Scheduler Selection in Locations With Significant Obstacles

Paulo Furtado Correia, André Coelho, Manuel Ricardo
INESC TEC, Faculdade de Engenharia, Universidade do Porto, Portugal
{paulo.j.correia, andre.f.coelho, manuel.ricardo}@inesctec.pt

*Abstract*— Integrated Access and Backhaul (IAB) in cellular networks combines access and backhaul within a wireless infrastructure reducing reliance on fibre-based backhaul. This enables flexible and more cost-effective network expansion, especially in hard-to-reach areas. Positioning a mobile IAB node (MIAB) in a seaport environment, in order to ensure on-demand, resilient wireless connectivity, presents unique challenges due to the high density of User Equipments (UEs) and potential shadowing effects caused by obstacles. This paper addresses the problem of positioning MIABs within areas containing UEs, fixed IAB donors (FIABs), and obstacles. Our approach considers user associations and different types of scheduling, ensuring MIABs and FIABs meet the capacity requirements of a special team of served UEs, while not exceeding backhaul capacity. With a Genetic Algorithm solver, we achieve capacity improvement gains, by up to 200% for the 90$^{th}$ percentile, particularly during emergency capacity demands.

## I. INTRODUCTION

Key societal sectors are becoming increasingly digitalised and data-driven. Sixth-generation (6G) networks aim at bridging the digital and physical worlds, providing resilient wireless connectivity to diverse verticals, including critical sectors such as seaports. These industrial environments have to enable communications between human operators, machinery equipment, sensors and video cameras, throughout their operational areas. In a seaport, there are terrestrial and nearshore areas, where operations take place. Container yards handle vast metallic loads with ships, cranes, and trucks in an increasingly automated environment. A 6G wireless network with properly positioned mobile communications cells offers a promising solution to meet the high capacity demanded by port users. For that purpose, these environments can take advantage of Mobile IAB nodes (MIABs) which are mobile gNodeBs carried by terrestrial vehicles or Unmanned Aerial Vehicles (UAVs) equipped with wireless backhaul, deployed on-demand to provide radio connectivity to User Equipments (UEs). They can cover obstructed areas improving wireless capacity to rescue teams in emergency or ongoing civil works.

In this paper, we investigate the added value of deploying a MIAB in a container yard at a seaport. We assume the following: 1) the seaport is covered by a set of fixed IAB Donors (FIABs) where obstacles, such as piles of containers, may severely attenuate the radio signal in certain areas; 2) the MIAB's backhaul link is wirelessly established via the NR Uu interface to a FIAB; 3) the MIAB is carried by a terrestrial vehicle, which may move throughout the seaport; 4) the MIAB aims to improve the communications provided to a specific group of users, called the special team. The focus is on the positioning of the MIAB within the port and exclude networking architectural aspects, which may benefit from the latest solutions proposed by 3GPP [19]. Obstacle-aware radio coverage is supplied by MIABs to maximise network capacity, thereby enhancing communications performance and improving the overall productivity of services within smart ports [1], such as emergency assistance truck rolls by special teams of users within individual areas.

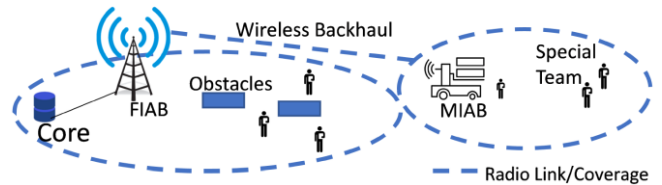

Fig. 1 – Concept diagram of a FIAB connected to a MIAB via wireless backhaul, jointly providing coverage to a set of UEs while accounting for potential shadowing caused by obstacles.

Our main original contributions include:

*1) A model for positioning MIABs and selecting the access and backhaul scheduler, in a 3D obstacle-aware environment:* we formulate the problem of positioning the MIABs considering the presence of FIABs, significant obstacles, areas where the MIAB can move, and a set of users requiring improved network performance. Additionally, the model considers two radio resource scheduling techniques - Round Robin (RR) and Proportional Fair (PF) - to determine the position for the MIABs that maximise the aggregate network capacity offered to a special team. For instance, commanders of first response team or personnel or equipment involved in a localised, temporary rescue or civil work.

*2) Network performance evaluation:* we evaluate the performance of the access and backhaul wireless network considering randomly generated scenarios at Terminal XXI, Port of Sines in Portugal, as shown in Fig. 1. The port is divided into well-delimited geographical areas. The remainder of this paper is structured as follows. Section II reviews the related work on positioning MIABs in outdoor areas, considering obstacles affecting the access and backhaul, cell selection, and scheduling algorithms. Section III describes the system model, including the main constraints, and performance metrics considered. Section IV presents the performance evaluation study and discusses the results obtained. Finally, Section V summarises the key findings and proposes directions for future work.

This work is co-financed by Component 5 – Capitalisation and Business Innovation integrated in the Resilience Dimension of the Recovery and Resilience Plan within the scope of the Recovery and Resilience Mechanism (MRR) of the European Union (EU), framed in the Next Generation EU, for the period 2021 – 2026, within project NEXUS, with reference 53.

## II. RELATED LITERATURE

Operations in seaports involving containers, trucks, ships, or cranes can cause shadowing, degrading the performance of legacy networks. In [3], wireless link disruptions caused by obstacles are mitigated through proactive handovers by considering links' history and their disruption probabilities. In [4], for the same type of mitigation, blockages are predicted using Deep-Learning. Both solutions are non-deterministic. Computer vision can also predict obstacles, users and their velocities, aiding in timely handovers as in [5]-[6]; however, these solutions have low accuracy in non-line-of-sight (NLoS) and low-luminosity conditions. In [7], computer vision aided by Machine Learning creates bounding boxes of obstacles to predict imminent blockages of wireless links; the accuracy of this solution is limited by the video cameras' resolution and the distances between objects. In [8], objects are modelled by cuboids and the NLoS conditions deterministically calculated using geometrical analysis; however, obstacles are reduced to zero-volume single points. Regarding the backhaul link, multiple solutions are provided in the literature. In [9], a solution is proposed for positioning of a mobile cell deployed in a UAV, considering also terrestrial base stations; this solution maximises the sum-rate but ignores obstacle shadowing in the backhaul link. [10] proposes a solution for drone-cells positioning, considering the constraints of backhaul capacity; however, it maximises the number of covered users rather than their sum-capacity. [11] considers only the backhaul link to maximise total network capacity for UAV base station positioning and channel allocation. [12] addresses the same problem but considers a free-space optical link for the backhaul; still, none of these solutions consider obstacles. [13] maximises the downlink sum-rate by controlling radio resources, transmission powers and UAV placement; this solution is backhaul-aware but still do not consider obstacle blockages. Optimal placement of base stations, given UEs' coordinates, is also considered in [14] for the minimum network cost conditions; however, it ignores obstacles and backhaul requirements. The literature provides solutions for base station positioning aimed at optimising various criteria, considering factors such as obstacle shadowing, cell selection procedures, backhaul capacity, resource allocation and scheduling algorithms. To the best of our knowledge, no published work addresses non-mmWave MIAB positioning to maximise network capacity while accounting for obstacles, backhaul capacity, cell selection, resource allocation, and scheduling.

## III. SYSTEM MODEL

Let us consider a three-dimensional networking scenario, as depicted in Fig. 2, where $M$ MIABs must be positioned within a given area of the seaport. The MIABs aim at providing wireless connectivity to a special team of $S$ UEs out of the total $U$, initially served by $F$ FIABs. The environment includes obstacles with substantial volumes that may disrupt the access or backhaul wireless links. The objective is to position the MIABs and determine their user associations with all UEs to maximise the aggregate network capacity offered to the special team within a given area.

Additionally, each MIAB's backhaul must support the capacity of the UEs it serves. In order to achieve this, the Euclidean coordinates for placing the MIABs ($x_m^M, y_m^M, z_m^M$) and the corresponding user associations ($S_{j,m}^M$) are determined to maximise the objective function.

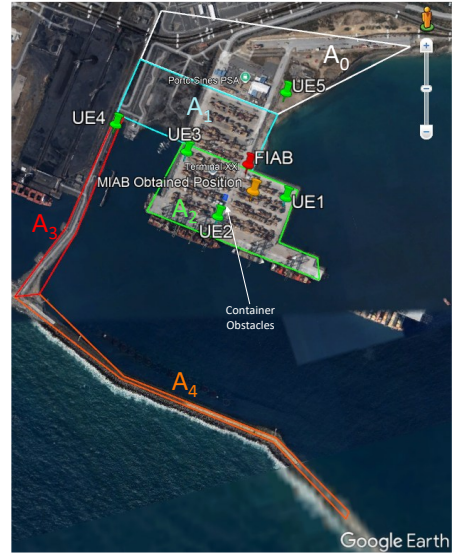

Fig. 2 - A top-down view of Sines Port's Terminal XXI in Portugal, with candidate deployment areas ($A_0$-$A_4$) for MIABs. FIAB (red marker) connected to a MIAB (orange marker) through its wireless backhaul, both providing coverage to UEs (*UE1-UE5*) located in well-defined areas subject to potential shadowing from significant obstacles.

### A. Problem Formulation

The problem is formulated as a Mixed-Integer Nonlinear Programming (MINLP) problem. The variables used to formulate the problem and their definitions are presented in Table I. The objective function in (1) maximises the sum of the network capacities offered to the $S$ UEs, ensuring single associations as denoted by (2). The 2D distances defined in (3), along with the applicable spectral efficiencies in (4), follow the models in [15]-[17]. For each access and backhaul wireless link, constraints (3) and (4) apply. The problem formulation is made generic, applicable to scenarios with $F$ FIABs and $M$ MIABs, each MIAB having a wireless backhaul to a single FIAB, as ensured by constraint (5). The aggregate capacity must satisfy constraint (7), where each MIAB's backhaul capacity ($C_{m,k}^F$) has to be at least the sum of the capacities offered to the UEs the MIAB serves. Moreover, each wireless link's Reference Signal Received Power (RSRP) must exceed a minimum threshold ($Qrxlevmin$), according to constraint (6), adapted from [2] to our scenario.

$$\max_{x_m^M, y_m^M, S_{u,m}^M, S_{u,k}^F, S_{m,k}^F} \sum_{u \in S} (\sum_{m \in M} C_{u,m}^M S_{u,m}^M + \sum_{k \in F} C_{u,k}^F S_{u,k}^F) \quad (1)$$

subject to:

$$\sum_m S_{u,m}^M + \sum_k S_{u,k}^F = 1, \forall u \in [1,U], \forall m \in [1,M], \forall k \in [1,F] \quad (2)$$

$$10 \leq d_{2Du,m}^M, d_{2Du,k}^F, d_{2Dm,k}^F \leq 5000 \quad (3)$$

$$0 \leq SE_{u,m}^M, SE_{u,k}^F, SE_{m,k}^F \leq 6.4 \quad (4)$$

$$\sum_k S_{m,k}^F = 1, \forall k \in [1,F] \quad (5)$$

$$R_{u,m}^M, R_{u,k}^F, R_{m,k}^F \geq Qrxlevmin \quad (6)$$

$$C_{m,k}^F \geq \sum_{u \in U} (C_{u,m}^M S_{u,m}^M), \forall m \in [1,M] \quad (7)$$

$$x_m^M, y_m^M \in A_a, z_m^M = h_{BSm} = 5 \quad (8)$$

TABLE I.  LIST OF THE MAIN NOTATIONS USED IN THE PROBLEM FORMULATION.

| Notation | Definition |
|---|---|
| $a, u, m, k$ | Indices representing area $A$, UE, MIAB, FIAB. |
| $MIAB_m$ | MIAB $m$, where $m$ ranges from 1 to $M$. |
| $FIAB_k$ | Fixed IAB donor $k$, where $k$ ranges from 1 to $F$. |
| $UE_u$ | User equipment $u$, where $u$ ranges from 1 to $U$. |
| $S$ | Number of UEs of the special team ($S \leq U$). |
| $A_a$ | The selected area $A$, where the special team requires capacity, as depicted in Fig.1. Each area is bounded by coloured lines defined by linear inequations as per the real scenario in Fig. 2. |
| $C_{u,m}^M, C_{u,k}^F, C_{m,k}^F$ | Access capacities provided respectively by $MIAB_m$ or $FIAB_k$ to the $UE_u$, and $FIAB_k$ to $MIAB_m$'s backhaul capacity, measured in bit/s. |
| $B$ | Resource Blocks (RBs) used per numerology. |
| $\Delta fr1\_m, \Delta fr1\_k$ | Subcarrier spacing (SCS) for $MIAB_m$ and $FIAB_k$ multiplied by 9 subcarriers per RB, in Hz. |
| $U_m, U_k$ | Number of UEs associated to $MIAB_m$ or $FIAB_k$. |
| $U_z$ | Number of RNTIs considered for RBs split, depending on the used scheduling type. |
| $S_{u,m}^M, S_{u,k}^F, S_{m,k}^F$ | Binary variables indicating user association of $UE_u$ to $MIAB_m$, $FIAB_k$, and $MIAB_m$'s backhaul association to $FIAB_k$ (1 if associated, 0 if not). |
| $Bw_{u,m}^M, Bw_{u,k}^F$ | Bandwidth of the channel respectively between $UE_u$ and $MIAB_m$ or $FIAB_k$, in Hz. |
| $Bw_{m,k}^F$ | Bandwidth of the backhaul between $MIAB_m$ and $FIAB_k$, in Hz. |
| $SINR_{u,m}^M, SINR_{u,k}^F$ | SINR of the channel respectively between $UE_u$ and $MIAB_m$ or $FIAB_k$, in dB. |
| $SINR_{m,k}^F$ | Backhaul SINR between $MIAB_m$, $FIAB_k$ in dB. |
| $R_{u,m}^M, R_{u,k}^F, R_{m,k}^F$ | Reference Signal Received Power (RSRP) by $UE_u$ respectively from $MIAB_m$ or $FIAB_k$, and by $MIAB_m$'s backhaul from $FIAB_k$ in dB. |
| $N_{fr1\_m}, N_{fr1\_k}$ | Thermal noise power for $MIAB_m$, $FIAB_k$ in dB. |
| $P_t$ | Transmission power in dBm. |
| $c$ | Speed of light in vacuum ($3 \times 10^8$ m/s). |
| $f_m^M, f_k^F$ | $MIAB_m$'s and $FIAB_k$'s carrier frequency in GHz. |
| $x_m^M, y_m^M, z_m^M$ | $MIAB_m$'s Euclidean coordinates in m. |
| $x_u^U, y_u^U, z_u^U$ | $UE_u$'s Euclidean coordinates in m. |
| $h_{BSm}, h_{BSk}, h_{UT}$ | $MIAB_m$'s, $FIAB_k$'s and $UE_u$'s heights in m. $h_{BSm} = 5$ m; $h_{BSk} = 10$ m; $h_{UT} = 1.5$ m, as in [16]. |
| $d_{2Du,m}^M, d_{3Du,m}^M$ | 2D and 3D Euclidean distances between $UE_u$ and $MIAB_m$ in m. |
| $d_{BPu,m}^M$ | 2D Euclidean breakpoint distances between $UE_u$ and $MIAB_m$ in m. |
| $\mu_{fr1\_m}, \mu_{fr1\_k}$ | FR1's numerology used by $MIAB_m$ or $FIAB_k$. |
| $\eta_{u,m}^M, \eta_{u,k}^F, \eta_{m,k}^F$ | Spectral efficiency of the established downlink RNTI respectively between $UE_u$ and $MIAB_m$ or $FIAB_k$, or between $MIAB_m$ and $FIAB_k$ in bit/s/Hz. |
| $SE_{u,m}^M, SE_{u,k}^F, SE_{m,k}^F$ | Obtained regression for spectral efficiency of the established downlink RNTI between $UE_u$, $MIAB_m$ or $FIAB_k$, as per (18), a linear function of the respective $SINR$, in bit/s/Hz, [18]. |
| $PL_{UMi\_LOSu,m}^M,$ $PL_{UMi\_NLOSu,m}^M$ | Path losses for UMi scenarios considered for LoS or NLoS links between $UE_u$ and $MIAB_m$. |
| $Qrxlevmin$ | Minimum RSRP needed to provide positive capacity, adapted to current configurations (-122 dBm) [2]. |
| $RB_{u,k}, RB_{u,m}, RB_{m,k}$ | Number of RBs respectively between $UE_u$ and $FIAB_k$, $MIAB_m$, and between $MIAB_m$ and $FIAB_k$. |

The problem solver provides the following results: 1) the aggregate capacity available to the special team of $S$ UEs; 2) the user associations of all UEs to the cells, represented by binary variables indicating each UE's associated cell; and 3) the MIAB's Euclidean coordinates adhering to (8). The capacities of the Radio Network Temporary Identifier (RNTI) between $UE_u$ and either $MIAB_m$ or $FIAB_k$ are defined in (9). They depend on several factors, including: 1) the spectral efficiency, defined in (10); 2) SCS multiplied by 9 subcarriers per RB, as defined in (11) for both frequencies; 3) the total number of RBs available per slot, and the number of UEs associated to a cell, which are allocated to connected UEs and backhauls; and 4) the scheduling type, PF or RR. PF allocates the RBs fairly among the $U_k$ UEs connected directly, as well as the $U_m$ UEs connected indirectly via $MIAB_m$'s backhaul, as specified in (12). RR treats the $MIAB_m$'s backhaul as a single UE and allocates the RBs equally based on the number of wireless links to $FIAB_k$. The allocation depends on the sum of backhaul associations $S_{m,k}^F$ and the $U_k$ UEs, as represented in (13).

$$C_{u,m}^M = \frac{B\ \Delta fr1\_m\ \eta_{u,m}^M}{U_m} \ ;\ C_{u,k}^F = \frac{B\ \Delta fr1\_k\ \eta_{u,k}^F}{U_z} ; \quad (9)$$

$$C_{m,k}^F = \begin{cases} \frac{B\ U_m\ \Delta fr1\_k\ \eta_{m,k}^F}{U_z} \ ;\ PF\ scheduler \\ \frac{B\ \Delta fr1\_k\ \eta_{m,k}^F}{U_z} \ ;\ RR\ scheduler \end{cases} \quad (9)$$

$$\eta_{u,m}^M = \begin{cases} SE_{u,m}^M\ ,\ SE_{u,m}^M \geq 0 \\ 0\ ,\ SE_{u,m}^M < 0 \end{cases} ;\ \eta_{u,k}^F = \begin{cases} SE_{u,k}^F\ ,\ SE_{u,k}^F \geq 0 \\ 0\ ,\ SE_{u,k}^F < 0 \end{cases} \quad (10)$$

$$\Delta fr1\_m = 2^{\mu_{fr1\_m}}.15kHz\ .9\ ;\ \Delta fr1\_k = 2^{\mu_{fr1\_k}}.15kHz\ .9 \quad (11)$$

$$U_m = \sum_{u \in U} S_{u,m}^M\ ;\ U_k = \sum_{u \in U} S_{u,k}^F \quad (12)$$

$$U_z = \begin{cases} U_k + U_m\ ;\ PF\ scheduler \\ U_k + \sum_{m \in M} S_{m,k}^F\ ;\ RR\ scheduler \end{cases} \quad (13)$$

Hence, the number of RBs assigned for each access link to $FIAB_k$ or $MIAB_m$ is respectively given by (14).

$$RB_{u,k} = \frac{B}{U_z}\ ;\ RB_{u,m} = \frac{B}{U_m} \quad (14)$$

Moreover, the number of RBs assigned to $MIAB_m$'s backhaul is determined by (15), depending on the scheduling type considered.

$$RB_{m,k} = \begin{cases} \frac{B\ U_m}{U_z}\ ;\ PF\ scheduler \\ \frac{B}{U_z}\ ;\ RR\ scheduler \end{cases} \quad (15)$$

The bandwidth used per RNTI for the direct wireless link between $UE_u$ and either $MIAB_m$ or $FIAB_k$ is given by (16). The bandwidth used by the backhaul of $MIAB_m$ is given by (17).

$$Bw_{u,m}^M = RB_{u,m}\ \Delta fr1\_m\ ;\ Bw_{u,k}^F = RB_{u,k}\ \Delta fr1\_k \quad (16)$$

$$Bw_{m,k}^F = RB_{m,k}\ \Delta fr1\_k \quad (17)$$

Equation (18) defines the spectral efficiency as a linear regression function of $SINR_{u,m}^M$, introduced, calibrated and demonstrated in [18], and validated in ns-3, as detailed in [17].

$$SE_{u,k}^F = 0.23\ SINR_{u,k}^F - 0.21 \quad (18)$$

The same linear regression model is applied to $SE^M_{u,m}$ and $SE^F_{m,k}$, which represent the spectral efficiencies for the UEs and backhauls of $MIAB_m$, respectively, as a function of their corresponding SINRs. This assumes the same MCS index table configuration is used for PDSCH, including modulation order, target code rate and corresponding spectral efficiency, as outlined in [15]. Applying (18) analogously in (10), results in spectral efficiencies $\eta^M_{u,m}$, $\eta^F_{u,k}$ and $\eta^F_{m,k}$, all of which positive.

The SINRs for each RNTI and the backhaul, in the absence of interference in the current model, are given by (19) and (20).

$$SINR^M_{u,m}=10log_{10}(\frac{R^M_{u,m}}{N_{fr1\_m}}); SINR^F_{u,k}=10log_{10}(\frac{R^F_{u,k}}{N_{fr1\_k}}) \quad (19)$$

$$SINR^F_{m,k} = 10\ log_{10}(\frac{R^F_{m,k}}{N_{fr1\_k}}) \quad (20)$$

The received power and thermal noise power for each RNTI and the backhaul are respectively provided by (21), (22) and (23), depending on LoS or NLoS conditions, which are determined by obstacle-aware modelling. In (21), $U_m$ is replaced by $U_z$ for the analogous $R^F_{u,k}$ and $R^F_{m,k}$ equations, depending on the type of scheduling used, as per (13).

$$R^M_{u,m}=\begin{cases} \frac{10^{(\frac{Pt-30}{10})}}{U_m\ PL^M_{UMi\_LoSu,m}} \ ;\ if\ LoS \\ \frac{10^{(\frac{Pt-30}{10})}}{U_m\ PL^M_{UMi\_NLoSu,m}} \ ;\ if\ NLoS \end{cases} \quad (21)$$

$$N_{fr1\_m}=10^{[-19.9+log_{10}(B\ \Delta fr1\_m)]} \quad (22)$$
$$N_{fr1\_k}=10^{[-19.9+log_{10}(B\ \Delta fr1\_k)]} \quad (23)$$

The RSRP in (21) relies on the 3GPP path loss model for a UMi scenario. In this model, $FIAB_k$ are all positioned at a height of 10 m, while $MIAB_m$ at a height of 5 m, as specified in [16]. As such, the problem becomes a true 3D optimisation, considering different heights among cells, UEs and obstacles. The path loss depends on (24) and (25). Equation (24), applied to LoS conditions, represents a two-branch function with a non-linearity point at a 2D breakpoint distance between UEs and cells. For NLoS conditions, it equals the maximum of (24) and (25), as shown by (26). Similar path loss expressions can be derived for $PL^F_{UMi\_LOSu,k}$, $PL^F_{UMi\_LOSm,k}$, $PL^F_{UMi\_NLOSu,k}$ and $PL^F_{UMi\_NLOSm,k}$.

$$PL^M_{UMi\_LOSu,m}=\begin{cases} PL^M_{1\ u,m},\ 10\ m \leq d^M_{2Du,m} \leq d^M_{BPu,m} \\ PL^M_{2\ u,m},\ d^M_{BPu,m} \leq d^M_{2Du,m} \leq 5\ km \end{cases} \quad (24)$$

$$PL^{M\ '}_{UMi\_NLOSu,m} = 10^{2.24}\frac{d^{M\ 3.53}_{3Du,m}f^{M\ 2.13}_m}{10^{(0.03(hUT-1.5))}} \quad (25)$$

$$PL^M_{UMi\_NLOSu,m}= max(PL^M_{UMi\_LOSu,m}, PL^{M\ '}_{UMi\_NLOSu,m}) \quad (26)$$

According to the definition in [16], $PL^M_{UMi\_LOSu,m}$, as in (24), is represented by $PL^M_{1\ u,m}$ for 2D distances below the breakpoint distance and takes the values of $PL^M_{2\ u,m}$ for 2D distances above, as presented in (27) and (28), respectively.

$$PL^M_{1\ u,m} = 10^{3.24} d^{M\ 2.1}_{3Du,m} f^{M\ 2}_m \quad (27)$$

$$PL^M_{2\ u,m} = 10^{3.24}\frac{d^{M\ 4}_{3Du,m} f^{M\ 2}_m}{(d^{M\ 2}_{BPu,m}+(hBS-hUT)^2)^{0.95}} \quad (28)$$

In UMi scenarios, the 2D breakpoint distance between a transmitter and receiver is a function of their antenna heights, the probability of LoS, the effective environment height ($h_E$) and the carrier frequency used, as shown in (31). $h_E$ is set to 1 m, with LoS probability of $1/(1+C(d^M_{2Du,m},h_{UT}))$, which equals 1 for UEs' heights ($h_{UT}$) up to 13 m.

The model excludes path loss values for 2D distances greater than 5000 m, which are beyond the range considered for $V$, and less than 10 m, as ensured by optimisation constraint (3). The distances $d^M_{3Du,m}$, $d^M_{2Du,m}$ and $d^M_{BPu,m}$ are given by (29), (30) and (31), respectively. Analogous equations can be inferred for distances involving $FIAB_k$, for accesses and backhaul, where the Euclidean coordinates, heights and frequencies of the parties involved are applied.

$$d^M_{3Du,m}= \sqrt{(x^M_m-x^U_u)^2+(y^M_m-y^U_u)^2+(z^M_m-z^U_u)^2} \quad (29)$$

$$d^M_{2Du,m}= \sqrt{(x^M_m-x^U_u)^2+(y^M_m-y^U_u)^2} \quad (30)$$

$$d^M_{BPu,m}= \frac{4\ (h_{BS}-h_E)(h_{UT}-h_E) f^M_m\ 10^9}{c} \quad (31)$$

In NR, the SCS for FR1 for the numerology $\mu_{fr1}$ = 1 result in a bandwidth of 30 kHz per RB. With 9 sub-carriers per RB and 133 RBs per slot, considering numerology 1, the total bandwidth available for data transmission is calculated as 133 x 9 x 30 kHz = 35.9 MHz. The binary variables $S^M_{u,m}$, $S^F_{u,k}$ and $S^F_{m,k}$, in constraints (2) and (5), ensure single $UE_u$ or backhaul associations with each cell. Constraint (3) ensures a lower bound 2D distance of 10 m respectively between $UE_u$ and $MIAB_m$ or $FIAB_k$, and the backhaul, so the 3GPP path loss $PL_1$ applicability range in [16] is valid. Constraint (4) bounds the spectral efficiency, as defined in [17], based on Table 2 of EESM NR error model type with Chase Combining. In the frequency domain, we use 9 sub-carriers per RB instead of 12, while in the time domain we use 12 OFDM symbols per slot, rather than 14. This is to account for overhead. Therefore, the upper bound of the spectral efficiency is 6.4 (bit/s/Hz) instead of 7.4 (for MCS index 27 and modulation order 8), according to [15] and [17].

*B. Obstacle-aware modelling*

In order to solve the MIABs' positioning problem, we consider obstacles such as containers, trucks, cranes or ships. For that purpose, we model obstacles as trapezoidal cuboids, with six planar faces and eight vertices defined by their Euclidean coordinates. The procedure deterministically and geometrically identifies any intersection point between one or more faces of an obstacle and the line connecting any $UE_u$ to either $MIAB_m$ or $FIAB_k$, or between the two connected by a wireless backhaul. The positions of UEs, MIABs or FIABs are identified by their Euclidean coordinates, while the obstacles' faces are defined by their vertices' coordinates and respective connecting lines' distances. For simplicity, the calculations to determine LoS or NLoS conditions are not presented herein. Moreover, NLoS conditions do not account for the specific attenuation introduced by the obstacle material. Hence, for any line considered, zero (LoS) or more (NLoS) intersection point(s) with face(s) of the obstacle(s) are identified, allowing for using respectively (24) or (26), to be applied in (21).

IV. PERFORMANCE EVALUATION

In order to evaluate the proposed model for positioning a MIAB in a 3D obstacle-aware environment, we defined a set of scenarios and evaluated the corresponding network performance by running a MATLAB script with the Genetic Algorithm (GA) solver. GA was chosen for its ability to explore large solution spaces and optimize MIAB positioning under a set of complex constraints; however, the proposed formulation is solver-agnostic and other methods may be explored. Table II provides the configuration parameter values used on the MATLAB and ns-3 simulations from [18]. After each run, GA provided several results, including: the aggregate capacity for the special team of UEs (UE1 and UE2), the gain compared with a baseline scenario, the user associations of UEs with the cells, the backhaul capacity of the MIAB, and the MIAB's position. Fig. 3 details the used methodology with variants *V0* to *V5* and the evaluated performance metrics. The baseline variant *V0* consisted of five UEs, all associated to FIAB. Two of these formed the special team, within each area $A_a$, while the other three could be in any area. The special team and FIAB were randomly positioned in each area $A_a$. *V0* has a single FIAB and no MIAB. We also defined five variants of the *V0* baseline – *V1* to *V5*, with a MIAB to assist FIAB. In *V1*, we added obstacles to *V0*; in *V2* and *V3*, we used PF scheduling, respectively without and with obstacles; in *V4* and *V5*, we used RR scheduling, respectively without and with obstacles. FIAB remained static for all scenarios in each area. For each of the five $A_a$ shown in Fig. 2, five different scenarios were defined for each variant (*V1* to *V5*), resulting in a total of 25 simulations per area, considering the methodology shown in Fig. 3. After simulating five different scenarios, for each of the six variants (*V0* to *V5*) for the five areas ($A_0$ to $A_4$), we obtained 5 x 6 x 5 =150 sets of results, including the baseline (*V0*). The objective was to compare the results obtained from *V0* (without MIAB), with the other five different variants (with MIAB). The gain of the special team aggregate network capacity for each variant, expressed as a percentage of the aggregate network capacity of *V0*, was calculated as ($C_{Vx}$-$C_{V0}$)x100/$C_{V0}$. The solver provided the aggregate network capacity of the special team, the gain percentage compared to *V0*, each $S_{u,m}^{M}$ or $S_{u,k}^{F}$, the backhaul capacity and the MIAB's position within each area. We then assessed the gains obtained for the aggregate network capacity of the special team as a function of the average topology distance (on the x-axis) and the inter-distance depicted by the red line (on the right y-axis), as depicted in Fig. 4 for area $A_0$. The average topology distance is the average distance between the UEs and FIAB. The inter-distance is the distance between the two UEs of the special team. Light blue bars show negative gains compared to *V0*, due to higher path loss attenuations due to obstacle shadowing. Green bars, representing PF scheduling, show higher gains without obstacles, as expected. Similarly, brown bars, representing RR scheduling, show the highest gains, especially without obstacles, for greater distances. This is due to always higher availability of RBs for the special team, despite being limited by the backhaul, combined with higher spectral efficiencies but with the appropriate user associations. Moving right on Fig. 4, higher separation between network nodes results in even higher gains. By focusing on the gains (125 results, from *V1* to *V5*) compared to *V0*, Fig. 5 depicts the cumulative distribution function (CDF) of each variant, for all the five scenarios of all the five areas studied. The light blue CDF confirms that obstacle shadowing typically causes a 50% reduction of the objective capacity. In most cases, RR is preferable to PF scheduling, except when obstacles are present (*V3* and *V5*), where performance is similar. Nonetheless, the MIAB consistently introduces gains in the objective function, increasing the aggregate network capacity for the special team, even in the presence of obstacles. The solver also calculated the backhaul capacity required to accommodate the combined network capacities of all the UEs directly served by the MIAB, which might not necessarily have been part of the special team. This confirms that the cell selection was adequately chosen by the solver to maximise the aggregate network capacities of the special team.

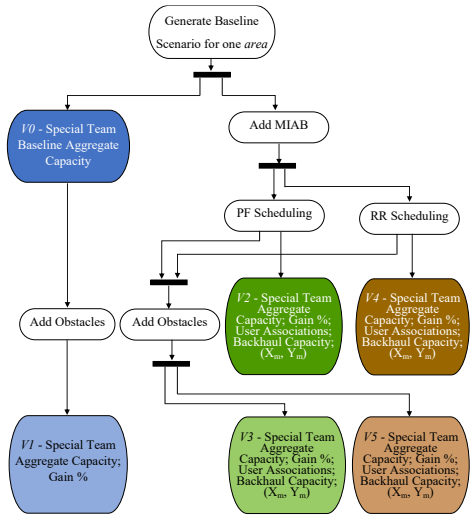

Fig. 3 – Methodology considered for network performance evaluation at each scenario. The coloured boxes summarise the results obtained for each variant (*V0* to *V5*).

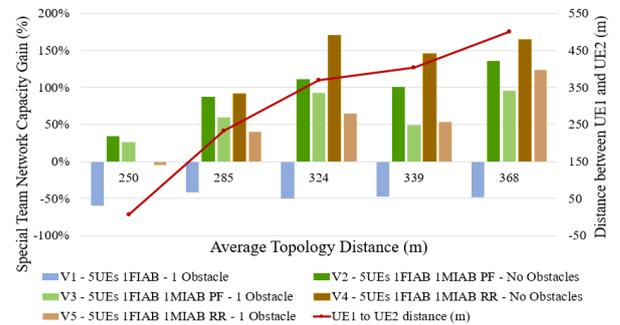

Fig. 4 - Aggregate network capacity gain percentages compared to the baseline *V0*, for the runs on $A_0$.

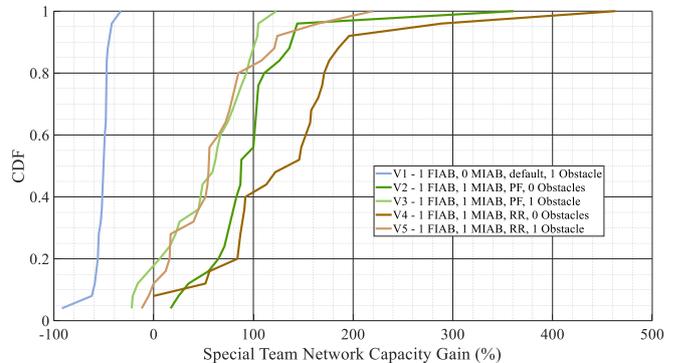

Fig. 5 - CDF of the aggregate network capacity gains for the special team.

Considering the backhaul capacity versus the aggregate capacity of the UEs served by the MIAB, Fig. 6 shows that the backhaul always supported the aggregate capacity for the UEs served by the MIAB, regardless of the variant.

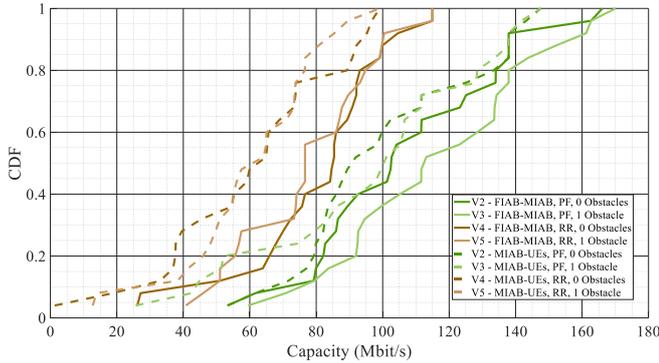

Fig. 6 - CDF of the aggregate network capacity provided to the UEs served by the MIAB versus its backhaul capacity from the FIAB.

TABLE II. PARAMETERS USED IN ALL THE RUNS OF THE SOLVER.

| Parameter | Value |
|---|---|
| 3GPP channel model | Urban Micro (UMi) |
| Number of MIABs and FIABs | 1 and 1 |
| Number of UEs and special team | 5 and 2 (UE1 and UE2) |
| Number of obstacles per scenario | 1 (randomly positioned) |
| MIAB's carrier frequency (GHz) | 3.9 |
| FIABs' carrier frequency (GHz) | 3.8 |
| Numerology ($\mu_{fr1\_m}, \mu_{fr1\_k}$) | 1 |
| SCS (kHz) | 30 |
| Antenna model | Isotropic (SISO) |
| Transmission Power (dBm) | 24 |
| Bandwidth (MHz) | 35.9 |
| Thermal Noise PSD (dB) | -123.4 |
| Traffic direction | Downlink |
| EESM Error Model | AMC with NrEesmCcT2 |
| NR MAC Scheduler Types | OFDMA PF or RR |
| Number of RBs per slot | 133 |
| Number of Subcarriers per RB | 9 |
| Number of OFDMA symbols per slot | 12 |
| GA (Population Size, Mutation Rate, Crossover rate) | (50, 20%, 80%) = MATLAB defaults / Fastest convergence |

## V. CONCLUSIONS

This paper proposes a system model formulated as an optimisation problem to determine the position of a MIAB including a performance evaluation study. The obtained MIAB's position maximises the aggregate capacity for a special team of UEs requiring improved network capacity within defined areas affected by significant obstacles. Our approach considers cell selection, the MIAB's backhaul capacity and two types of scheduling for each scenario: PF and RR. The proposed solution, designed for deployment at the MIAB, allows for gains up to 200% for the 90[th] percentile, especially when using RR scheduling, while addressing obstacle-induced shadowing. Moreover, the backhaul wireless link always supports the capacity provided by the MIAB to the served UEs. Through these contributions, we aim to support Mobile Network Operators in optimising IAB networks planning for outdoor environments, such as seaports, typically characterised by multiple UEs, FIABs, MIABs, and significant obstacles. As future work, we will explore scalability with multiple MIABs and FIABs, and leverage O-RAN xApps for autonomous deployment optimization.